\documentclass[runningheads]{llncs}
\usepackage{graphicx}
\usepackage{amssymb, amsmath, amsfonts}
\usepackage{graphicx}
\usepackage{color}
\usepackage{mathrsfs}
\usepackage{physics}
\usepackage{mathtools}
\usepackage{tikz}
\usepackage{faktor}
\usepackage{tikz-cd}
\usepackage{bm}
\usepackage{hyperref}
\usepackage{csquotes}

\usepackage[UKenglish]{isodate}
\usepackage[UKenglish]{babel}

\DeclareMathOperator{\E}{\mathbf{E}}
\newcommand{\lag}{\mathcal{L}}

\begin{document}
\title{Resilience and adaptability in self-evidencing systems} 
\titlerunning{Resilience and adaptability in self-evidencing systems}
%

\author{Mahault Albarracin\inst{1, 2,}\orcidID{0000-0003-0916-4645} \and Dalton A R Sakthivadivel\inst{1,3,}\orcidID{0000-0002-7907-7611} }
%
%
\institute{VERSES Research Lab \and D\'epartement d'informatique, Universit\'e du Qu\'ebec \`a Montr\'eal, Montr\'eal, Canada \and Department of Mathematics, CUNY Graduate Centre, New York, NY 10016\\
\email{dsakthivadivel@gc.cuny.edu}}

\authorrunning{M Albarracin and D A R Sakthivadivel}

\maketitle              
\begin{abstract}

In this paper we will articulate a view of resilience under the free energy principle and {\it vice versa}. The free energy principle is about existence and identity, and resilience is the condition under which things exist at all. In previous work this has been investigated as modelling resilience using the free energy principle. We will extend that work by making the case that self-organisation under the free energy principle is {\it about} resilience, in the sense that identity is a constant process of self-reconfiguration, implying the existence of a self-model and the energy to reconfigure that self-model---and hence, the resilience of a maintained identity under changes. A general framework for thinking about resilience in this context will be sketched out and some models will be provided using that framework.

\end{abstract}
\section{Introduction}


This paper is mainly concerned with the notions of identity and resilience which are at stake when modelling adaptive systems. Self-organisation is implicitly defined in terms of identity---that is, the self. However, it is also apparent that self-organisation is never static, and the sense of identity implicit here must somehow be flexible or else be capable of accommodating changes in the environment. In dissipative adaptation, for example, a system self-organises by restructuring itself to absorb and dissipate energy from an external source in the environment \cite{england2015dissipative,horowitz2017spontaneous,kachman2017self,prigogine1978time}. This echoes enactive and autopoietic theories: living systems produce and maintain themselves \cite{maturana2012autopoiesis}, and for an organism to maintain a model of itself and its environment, it must produce what is required to preserve its generative model \cite{mazzaglia2022free}. Maturana--Varela's concept of autopoiesis describes how a system continuously regenerates its components but also keeps its identity intact---again encountering the tension we highlighted. This paper aims to investigate what an appropriate description of this flexibility might be and how can it can be modelled, and in particular, what resilience has to do with a changing identity in dynamic environments. We will develop a first idea of a formalism to describe resilience and some simple models for how resilience features in adaptive systems.

Adaptation is often achieved by, or can be modelled as, an inference process \cite{friston2015knowing,pachter2024entropy,perunov2016statistical,seifert2012stochastic,ueltzhoffer2021drive}. An interesting example to study is the human brain. Modern theoretical neuroscience asserts that the brain maintains an internal self-model which is continuously refined as part of its predictive engagement with the world. In predictive processing, an agent must model the causes of its sensory inputs to respond to perturbations from the environment \cite{friston2012prediction}. Since the agent's own body, traits, and actions are implicated in those causes, the brain encodes a model of `self' to better predict incoming sensations \cite{hohwy201716}. In particular, Hohwy--Michael suggest that people ``perceive and maintain their selfhood by an internal model of hierarchical endogenous (hidden) causes'', wherein higher-level causes like goals and traits generate predictions that shape perception and action. It follows that identity is embodied in the brain's generative model and is continuously updated by maximising model evidence and minimising prediction error, a process one might call {\it self-evidencing}. 

As discussed, maintaining an identity requires not just change but also a degree of stability of core features, offering a conundrum. To self-evidence, an agent learns and constantly adjusts a generative model of the world and of itself, using sensory evidence to correct discrepancies. The brain expends informational and energetic resources to keep its self-model aligned with reality---revising beliefs or deploying actions when mismatches (prediction errors) arise. This continual updating is the computational essence of self-reconfiguration. The self-model is never static, because new experiences demand incorporation.

One way to resolve this tension is to view an agent as balancing accuracy (updating to fit data) with precision or confidence in prior beliefs, as is done in active inference \cite{friston2017active,miller2022resilience}. 
Some self-model predictions---{\it e.g.} deeply held traits or values---are high-precision priors that resist change, providing continuity of identity. This balance is evident in psychology: people seek to preserve a coherent self-concept in the face of discrepant information (the motive for self-consistency or self-verification) even as they update other aspects of themselves over time \cite{mokady2022role}. For example, individuals continuously gather self-relevant information to build an inner `self-theory', updating beliefs about themselves with feedback from social and personal experiences.
Over time, as a more stable self-concept is consolidated, people begin to give greater weight to maintaining existing self-beliefs, ``at the expense of continuous updating''.

Altogether this implies that the ``self'' is not a static essence but a working model, continually regenerated by the very organism it defines. To characterise the malleability of certain aspects of identity and the rigidity of certain other ones, we will frame it in terms of {\it resilience}. An identity (the organism's self-model) is sustained by ongoing self-production and error-correction---the same processes that characterise  resilience. Resilience generally refers to the capacity of a system to withstand shocks or disturbances without losing its core functions or identity \cite{hanefeld2018towards,miller2022resilience,ramasubramanian2022influence}.

The term resilience is polysemous: it encompasses at least three complementary aspects \cite{albarracin2024sustainability,miller2022resilience}. It denotes (i) inertia---resistance to change when disturbed, (ii) elasticity---the ability to rebound or restore a prior state after a deviation, and (iii) plasticity---the capacity to flexibly reorganize or find new stable states when conditions change. All three aspects describe processes rather than static traits. Resilience is not simple a fixed quantity one has a certain amount of; it is something one does, dynamically, over time. 

The link between identity processes and resilience becomes apparent when we consider what is maintaining or recovering in each case. In the face of life's perturbations---be it trauma, loss, illness, or rapid change---it is often one's sense of self and purpose that is challenged. Awareness and cognitive flexibility distinguish the more resilient phenotypes, allowing agents to anticipate and integrate surprises without being overwhelmed. A resilient individual manages to hold onto an integrated self-concept (core values, sense of meaning, self-esteem) or to reconstruct a new one, such that they continue to function and grow. This points to multiple scales of the self, akin to Russian dolls with varying degrees of malleability---or rates of change. 

\begin{figure}
    \centering
    \includegraphics[width=\linewidth]{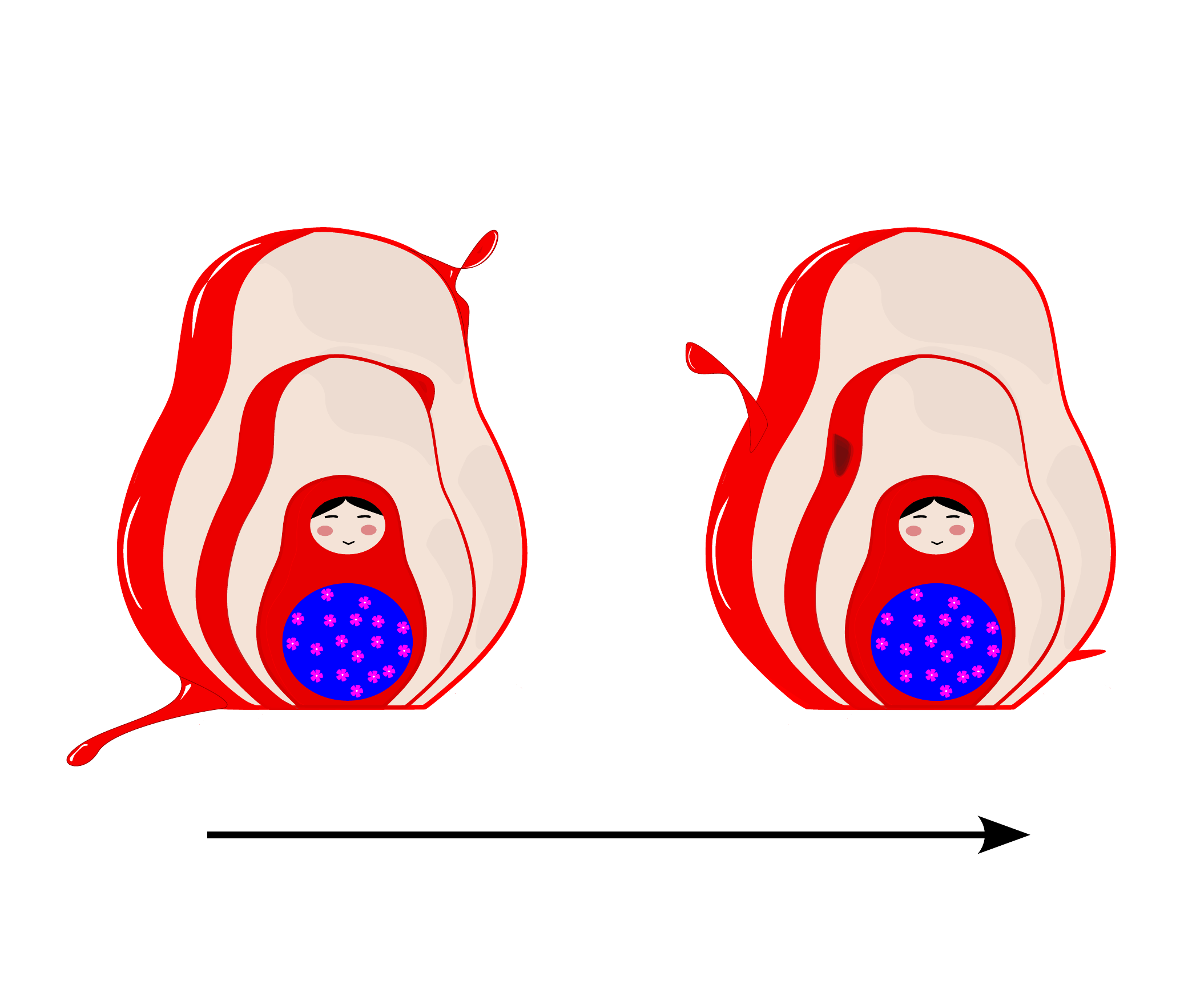}
    \caption{The Matryoshka doll exhibits a nested sense of identity. As time goes on, defects and mutations in the outer dolls change their phenotypes. The innermost doll, however, remains intact and constant.}
    \label{fig:dolls-fig}
\end{figure}

Identity has been linked in different ways to resilience. the concepts have porosity in the literature and a degree of metaphorical artistic license. Empirical studies show clear connections between certain qualities of identity and resilience outcomes. For instance, people who have more complex identities---{\it i.e.} who see themselves in terms of multiple distinct roles or attributes---tend to be more resilient to stress and negative events \cite{chen2012identities,shih2019multiple}. If one facet of the self is threatened, a person with many facets can draw strength and continue to find self-worth in other aspects, preventing a total collapse of self-regard. This flexibility in self-concept, a form of adaptive reconfiguration of identity, is a protective factor. Similarly, having a coherent but adaptable narrative identity (a life story that makes meaning of difficulties) is associated with resilience \cite{hall2016engendering,kerr2020complexity}. Individuals who can `re-story' their lives after trauma, incorporating the adversity into their identity as a survivor or as someone who has grown, often show better recovery and even post-traumatic growth \cite{meichenbaum2017resilience}. Conversely, a brittle and rigid identity will lead an individual to not be able to recover from any major disruption. 

Ultimately, it is the degree of continuity of identity which is paramount. Breakwell (2021) introduced the notion of identity resilience within identity process theory, defining it as ``a relatively stable self-schema based on self-esteem, self-efficacy, positive distinctiveness and continuity'' that helps one cope with threat \cite{breakwell2021identity}. In this sense, resilience can be seen as the successful management of identity threats: maintaining core identity integrity whilst adapting aspects that need to change.

Given the above, we can argue that the process of identity formation and reconfiguration is inherently a resilience process.

In that context, a resilient response corrects a disturbance or accommodates it by updating the self-model. This corresponds to elasticity and plasticity, with resilient elasticity being an agent's natural return to its characteristic states ({\it e.g.} attractors) after a shock, and resilient plasticity being the possession of functional redundancy or degeneracy--- that is, multiple ways to achieve well-being, or allowing flexible reorganisation. 

Resilience-as-inertia, meanwhile, corresponds to holding very high-precision beliefs that prevent change. In active inference, an agent with very precise self-beliefs might appear `resilient' in the sense of resisting influence (robust identity), but could also verge on inflexibility. An agent with more degenerate solutions (many possible adaptive states) can explore alternatives when one mode of being is compromised. In psychology, this sense of resilience naturally emerges from the concept of an enactive {\it niche}, the physical and social environment an individual helps shape around themselves \cite{harrison2025resilience}. An agent might proactively seek supportive relationships and stable routines (shaping their niche) to buffer against future stressors,  preserving their identity and well-being. The agent infers states of the world and selects actions that keep them within a regulated zone, and those actions simultaneously alter the environment in their favour.

Resilient adaptation can be seen as an ongoing negotiation of identity. The individual or system must decide what to hold onto and what to change in response to perturbation. Maintaining functional integrity might mean preserving certain core self-aspects (values, roles) constant (inertia), while adaptive reorganisation might mean transforming other aspects (adopting a new identity, goal, or belief system) to accommodate new realities (plasticity). 

\section{A basic mathematical formalism}

We can think of a system's identity configuration or psychological state as a point in a high-dimensional state space. 
Over time, as the system goes through typical experiences, their state traces out a trajectory---essentially their equilibrium or habitual behaviour. Resilience relates to what happens when the system is pushed off this usual trajectory. 


Certain configurations of the self may be stable attractors (basins of attraction) representing a well-integrated identity or mode of functioning. Perturbations might nudge the system within the basin (from which it can return to the attractor, recovery) or, if strong enough, knock the system into a different basin (a qualitative shift in identity or functioning). This is analogous to how ecosystems have multiple stable states; resilience is the difficulty of tipping the ecosystem into an alternate state. 

High-precision beliefs create deep attractors (inertia), whereas plasticity creates a flatter landscape with multiple smaller attractors, allowing movement to new states. We will focus on the high-precision case first.




\subsection{Inertia}

We will begin by supposing we have a (uniformly Lipschitz continuous) time-varying vector field $V$ and a random variable satisfying 
\begin{equation}\label{sde-eq}
\dd{X_t} = V(X_t, t) \dd{t} + \Gamma\dd{W_t},
\end{equation}
in other words, fluctuating around the expected path $\hat{x}_t \coloneqq \int_0^t V(X_s, s) \dd{s}$, with $\Gamma$ a positive-definite matrix. We will assume $\Gamma$ is independent of $X_t$ for convenience; the same story we will tell can be told for state-dependent noise, but it is considerably more complicated. Let $\abs{\Gamma} = \det \Gamma > 0$. By a theorem of Friedlin--Wentzell, the non-equilibrium measure over paths $p_{\abs{\Gamma}}(X_t)$ is given (to first order in $\abs{\Gamma}^{-1}$) by
\begin{equation}\label{fw-thm-eq}
e^{-\frac{1}{\abs{\Gamma}} \int^t_0 \abs{\dot{X}_s - V(X_s, s)}^2 \dd{s}}.
\end{equation}

Suppose $p(X_t)$ is the Bayesian belief of an agent about its own states. Then $\abs{\Gamma}^{-1}$ is the precision of that belief. When $\abs{\Gamma}^{-1}$ increases, the probability of large fluctuations away from $\hat{x}_t$ decays exponentially. That is to say, as $\abs{\Gamma}^{-1}$ increases, time evolutions $X_t$ that accumulate large distances from $\hat{x}_t$ in time are penalised more and $p(X_t)$ decreases. A formal computation reveals the following. In the limit, we have 
\[
p_\infty(X_t) = e^{-\infty \int^t_0 \abs{\dot{X}_s - V(X_s, s)}^2 \dd{s}} 
\]
which is only non-zero when 
\[
\int^t_0 \abs{\dot{X}_s - V(X_s, s)}^2 \dd{s}
\]
vanishes, in which case it is $e^{-\infty \cdot 0} = 1$. Another way to see this is as follows. As $\abs{\Gamma}^{-1} \to \infty$, we have $\abs{\Gamma} \to 0$, giving us a zero-noise limit. Then from \eqref{sde-eq} it is immediate that the system does not fluctuate away from the expected path. Since it is highly probable for the system to evolve near $\hat{x}_t$ when precision is high, it becomes a path-wise attractor. If the paths around the expected path are all orbits through the same region of the state-space, it moreover becomes an attractor in the state-space. As such, resilience in the inertial sense is high self-precision.


Following \cite{friston2023path,sakthivadivel2022worked} we will now articulate how this connects to the free energy principle. First we will rehearse the FEP according to those references.



Let $\lag$ be the log-probability. Suppose the system is coupled to some other system, such as an environment. The free energy principle begins from the observation that the KL divergence
\[
F(\mu_t, b_t) \coloneqq \int q(\eta_t; \sigma(\mu_t)) \log q(\eta_t; \sigma(\mu_t)) \dd{\eta_t} - \int q(\eta_t; \sigma(\mu_t)) \log p(\eta_t,  b_t, \mu_t) \dd{\eta_t}
\]
decomposes as 
\[
D_{\mathrm{KL}}(q(\eta_t; \sigma(\mu_t), p(\eta_t \mid b_t)) + \lag(\mu_t, b_t).
\]
Since the KL divergence is non-negative, we moreover have 
\[
D_{\mathrm{KL}}(q(\eta_t; \sigma(\mu_t), p(\eta_t \mid b_t)) + \lag(\mu_t, b_t) \geqslant \lag(\mu_t, b_t)
\]
saturating at the modal path, or the expected path under the Laplace assumption. A Taylor expansion arguments allows us to conclude 
\begin{equation}\label{abil-eq}
\nabla_{\mu_t} \lag(\mu_t, b_t) = 0 \iff \nabla_{\mu_t} F(\mu_t, b_t) = 0.
\end{equation}
Now we have licence to claim that the modal path encodes a variational inference about the environment: it is the state for which the Lagrangian is minimised, and hence, the parameter of the posterior distribution over external states for which the variational free energy is minimised. 

Conceptually this is a tautology. The free energy principle is a statement that {\it if} a system is on an attractor then it looks as if it is minimising a variational free energy and the dynamics on the attractor (as well as fluctuations around it) can be modelled using whatever dynamical system is induced by approximate Bayesian inference (regarded as a law of motion) \cite[\textsection 2]{ramstead2023bayesian}. The fact that this statement has a direction is crucial: we assume it is a `thing' at the outset. Once we have a thing, the FEP is a highly general method of modelling `things' and describing {\it why} they are things (in particular, the things that they are).

Now we will connect this to our argument. By \eqref{fw-thm-eq}, we have the {\it surprisal} or {\it rate function}
\[
-\log p_{\abs{\Gamma}}(\mu_t) = \frac{1}{\abs{\Gamma}} \int^t_0 \abs{\dot{\mu}_s - V(\mu_s, s)}^2 \dd{s} + o\left(\frac{1}{\abs{\Gamma}}\right).
\]
Minimising variational free energy also minimises this quantity by \eqref{abil-eq}. Consequently it increases precision. 


We see that as beliefs about the environment become better approximations of the distribution over environmental states, highly precise self-beliefs are afforded to the system, enabling it to remain close to characteristic trajectories through state space. In the other direction, having highly precise self-beliefs demands an accurate world model: if variational free energy cannot be minimised, then neither can the surprisal, and the system will drift from its characteristic trajectory.

This has a dual viewpoint under the principle of maximum entropy \cite{g-and-a}. A constraint imposes acceptability or unacceptability on different states of the system. It may imply the existence of a target morphology, or some desired form driven by a function---something we may call a phenotype. This ultimately coincides with what we have called an identity \cite{kiefer}.


\subsection{Plasticity}

The previous section assumed the system was capable of forming a sufficiently good model of the environment to minimise $D_{\mathrm{KL}}$. Recall that if variational free energy cannot be minimised, then neither can the surprisal, and the system will drift from its characteristic trajectory. As such, it must designate new trajectories as acceptable---its identity must be {\it plastic}. The notion of plasticity was introduced to describe systems that are able to grow and change their characteristic trajectory.




We will now work state-wise. Let us make the Laplace assumption such that to every possible posterior over environmental states there is a unique associated (mode, variance) pair. Suppose the statistics of the environment change. From \eqref{abil-eq} it follows that a new mode must be chosen by the system. It is possible for a system to adjust the statistics of the environment to bring the corresponding internal mode closer to a preference, but it is not always possible to exactly meet those preferences. 

\subsubsection{Changing modes}

Recall that the mode is actually the conditionally expected mode given the blanket state. In effect, there is an entire suite of modes, each a local minimum of the rate function on the joint state space offering its own phase. Said differently, these modes offer different attractors or metastable states. When the magnitude of a perturbation is too large, it may kick the system off of one attractor and it may be caught in the well of another. Whether one or another mode is acceptable depends on the statistics of the environment, so we have different phenotypes in different environmental parameter regimes, with the choice of mode determined by some control parameter---here, the blanket state. If we have a bidirectional coupling of the agent to its blanket, {\it e.g.} $\dd{\mu_t} = V(\mu_t, b_t, t) \dd{t} + \Gamma \dd{W_t}$, then the mode of the internal states is also a function of the external state {\it via} $\sigma$, and evolves along a trajectory going between such metastable states. (Note the internal states cannot see the external states directly, and any effect the environment has on the agent is mediated by the blanket.)

A physical example of the situation we have in mind is that of a phase transition. In a phase transition, the same Hamiltonian characterises a material, but the typical dynamics are very different depending on the value of some control parameter, and there is a sharp discontinuity between different phases. 



A simple model thereof is the Ising model on a lattice $\Lambda$ coupled to a heat reservoir. There is an internal state---the spin configuration $\ket{s_1s_2s_3\ldots s_\abs{\Lambda}}$---an external state---the state of the reservoir in the environment---and a blanket---the physical interface between the material and the environment, and the heat bath with temperature $T$ between the system and whatever environmental dynamics are taking place. Importantly, the heat bath is in contact with the material but shrouds the environment from it, serving as a Markov blanket. 

Assume there is some unknown but non-decreasing transfer function relating the environment to the heat bath. As the temperature of the environment increases, the temperature of the heat bath also increases. Now the mode of the system (its expected internal state) depends on $T$. The mode is the well-known magnetisation function $m(T)$. The mode encodes a posterior in the sense that if the heat bath is hot then the environment likely does not contain any heat sinks, and likely contains heat sources, or at least is in equilibrium with the heat bath. 

Though beyond the scope of this paper, it is interesting to note that large deviations interact with metastability and phase transitions, following general principles laid out in \cite{olivieri2005large}.

\subsubsection{Free energy curvatures}

We also would like to model more continuous changes in identity, not dependent on sharp discontinuities in a control parameter. If the blanket state is constant or slowly changing, another route to exploring different attractors and introducing local minima is decreasing the curvature of the free energy functional. By a Taylor expansion argument, the curvature of the free energy---the matrix of second derivatives of the KL divergence in the parameters---is the Fisher information.

This has the interpretation that when the curvature of the free energy is high in a certain direction, the system will find those directions more informative and travel in those directions. That is to say, the Fisher information metric measures how much information each parameter carries about distinguishing between nearby distributions. This is the opposite of intertial resilience in a precise sense: under the Laplace assumption, the Fisher information is proportional to the precision (see {\it e.g.} \cite{friston2007variational} for a lengthier discussion), so that higher precision makes for deeper curvatures, and hence stronger attractors. Indeed, for Gaussians, directions where the distribution has low variance---large eigenvalues of the precision matrix---provide more information for parameter estimation, whilst directions with high variance---small eigenvalues of the precision matrix---provide less information. By flattening the free energy landscape, the system is not offered strong information from parameters and hence has little preference for what variational posteriors to use, and is able to explore rather than being trapped in a deep global minimum. 


We claim that an adaptive system will find more effective ways to self-organise when the free energy landscape is flatter. To justify this, note that the heat dissipated is (see \cite{parr2020markov,seifert2012stochastic} for extended derivations)
\[
\bm{q} = -k_B T \Delta F(\mu, b)
\]
where $\Delta$ is the sum of unmixed second partials, or in other words, the trace of the Hessian matrix. 

If every change in $\mu$ is a new possible parameter, then the $\Delta F(\mu, b)$ term is the Fisher information. Since the trace of a square matrix is the sum of its eigenvalues, the heat dissipated is proportional to the negative precision. It follows that low precision dissipates a larger amount of heat than high precision, suggesting that when the curvature is small, self-organisation can be made more energetically-efficient. 

\section{A model of complex adaptive processes}


To recap, there are (so far) two different notions of resilience that have been explored. In \cite{miller2022resilience} these phenomena were distinguished as {\it inertia} and {\it plasticity}---the ability to resist change, and the ability to respond to change, respectively. In the previous subsection we discussed how a threshold for damage distinguishes between these forms of resilience, in the sense of the continuity and discontinuity of changes in configuration. There is something within active inference that makes a stone rigid and tough, and something that makes a human vulnerable but flexible. The task of this section is to find it.

One conjecture is that this threshold is an emergent effect of the distributed, hierarchical nature of a complex system. A human can modularise and spread damage amongst many lower level processes to protect a higher-level process like consciousness. Its identity is not stored in the configuration of a single level of resolution. For example, a human being bleeds from many different capillaries or arteries rather than a single heart. Contrast this with a cell, which has no such highways and when punctured dies immediately; or a honeybee, which has no blood vessels at all, and delivers oxygenated blood around its body by diffusion. (Note that this tacitly assumes hierarchy implies modularity.) A similar argument is made in \cite{sajid2020degeneracy} and is used in \cite{miller2022resilience}.

To that end we will discuss a hierarchical free-energy minimising system. Consider a system with a suite of $m$ internal states $\bm \mu \coloneqq \{\mu_1, \ldots, \mu_m\}$, $k$ active and $p$ sensory states $\{a_1, \ldots, a_k, s_1, \ldots, s_p\}$ (for a total of $k+p$ blanket states), situated in an environment with $n \ll m$ states $\bm\eta \coloneqq \{\eta_1, \ldots, \eta_n\}$. Each state will be given by a random variable evolving in time. 


Under generic conditions there exists a mapping from the conditionally expected trajectory of $\bm \mu$ given $\bm a, \bm s$ to that of $\bm \eta$,
\[
\sigma: \hat{\bm\mu}_{\bm a, \bm s} \mapsto \hat{\bm\eta}_{\bm a, \bm s}.
\]
This mapping serves to relate the parameters of a conditional probability distribution over external states to the expected state of the system, 
\[
P(\eta; \hat{\bm\eta}_{\bm a, \bm s}, \Sigma_{\bm a, \bm s}) = P(\eta; \sigma(\hat{\bm\mu}_{\bm a, \bm s}), \Sigma_{\bm a, \bm s})
\]
In our situation we have a redundancy in the parametrisation: some fraction of the information in each $\hat{\bm\eta}_i$ is mapped to each $\hat{\bm\mu_i}$. An example of this redundancy is a human brain, where many more sensory neurones exist than environmental features. 


Similarly, suppose the internal states are grouped into $v$ clusters, denoted $\bm C \coloneqq \{C_1, \ldots, C_v\}$, each of which has a mode given by the most likely state within the cluster. Suppose the mode of the cluster also parametrises some free energy functional over environmental states. This may be over coarse-grained features of the environment, or may simply be a cluster of repeated representations of the same feature. We will denote these external states as $\tilde\eta$. We will also assume there is a corresponding possible difference in the blanket states of interest, $\tilde b$.


Let the empirical mean of a cluster of $h$ states $C$ be
\[
\hat{C} = \frac{1}{h}\sum_i^h \mu_i
\]
and (under the Laplace assumption) let $\hat C$ parameterise a distribution over $\tilde\eta$ by 
\[
D_{\mathrm{KL}}(q(\tilde\eta; \tilde\sigma(C), p(\tilde\eta \mid \tilde b)) + \lag(C, \tilde b).
\]
All of our previous reasoning applies, so that the optimal state of the cluster is $\E[C]$. 

The probability of the sample mean of $h$ iid normally distributed random variables deviating from $\E[C]$ by $\varepsilon$ can be estimated as
\[
P(\hat{C} - \E[C] \geqslant h\varepsilon) \leqslant e^{-h\varepsilon^2}
\]
meaning that as $h$ grows large, $\hat{C}$ is exponentially likely to get closer and closer to $\E[C]$. Conceptually, as we have more and more states in the cluster, losing a state makes less of an impact on how far the average of the cluster is to the optimal parameter. That is to say: if some states are lost but there are sufficiently many states, the mode of the cluster does not change a large amount. As such, the higher level processes remain intact. 

Though an adequate first idea, a more sophisticated argument would use the relationship between metastability and multiscale large deviations estimates \cite{ellis1999theory,li2024large} to expand this framework. We leave this to future work.


\section{Closing remarks and a discussion of psychology}

We hope the framework sketched in this paper will provide more effective ways of modelling not only adaptive systems in physics but also the human brain and emotions. Like the discussion here, classic and contemporary psychology frames identity as an ongoing, adaptive process rather than a fixed entity. Symbolic interactionist theories ({\it e.g.} Mead’s and Cooley’s ideas of the self) emphasize that identity forms through social feedback and iterative self-reflection \cite{cooley1902looking,mead1913social}. A striking formalization comes from identity control theory in sociology, which models each identity ({\it e.g.} role or group identity) as a homeostatic feedback loop \cite{deane2021shape,ikegami2008homeostatic,seth2018being}.
An identity includes a mental standard (the meanings and expectations of being {\it e.g.} ``a good parent'' or ``a proud student''), and individuals continuously compare their perceived behaviour or feedback (input) to this internal standard (comparator) \cite{proust2003thinking,stets2003sociological}, which is ultimately a form of sociometry \cite{albarracin2024feeling}. 

If a discrepancy (error) is detected---say, others' reactions suggest one is not meeting the identity standard---the individual will adjust either their behaviour or their identity perceptions to reduce the error.
In effect, each identity acts as a control system. It is meant to control internal and external perceptions by making them congruent with the meanings of its identity standard. In this way, any error is minimized.
Identity is enacted and tuned in real time as situations change. The individual’s self-model (their identity standard and current self-perception) is thus iteratively revised to better match experience, or conversely, they act on the world to make experience match their identity \cite{morris2019fluid,woodward2018concepts}.

Identities can reorganize over longer timescales through processes like developmental crises or narrative reframing \cite{cote2018enduring,howse2024coping}. Erikson's classic stages of identity formation and modern notions of narrative identity both see identity as an evolving story, continuously edited to make sense of new chapters of life \cite{ergun2020identity,mcadams2011narrative}. For instance, Epstein’s cognitive-experiential self theory posits that people gradually construct a self-schema from life experiences, which serves as a working theory of who they are \cite{epstein1973self}.

Early in life this schema is highly malleable---young people update their self-concept rapidly with new feedback---but as it stabilizes, they increasingly strive to maintain coherence, sometimes by filtering or reinterpreting discrepant information rather than altering the self \cite{mokady2022role}. Consequently, a healthy identity involves both change and consistency, akin to a ship that adjusts course in changing seas while keeping a steady direction.
This dynamic is facilitated by cognitive capacities (memory, reflection) and informational resources (feedback from relationships, culture) that allow the person to revise their self-narrative when needed. It also has an energetic aspect: adapting one's identity---for instance, during acculturation or personal transformation---can be effortful and emotionally taxing, indicating that resilience of identity may depend on having adequate emotional and cognitive energy to accommodate change.

All of these observations suggest a place for resilience in the free energy principle---and more crucially, a reading of the free energy principle where resilience is a key concept underwriting self-assembly and self-organisation.

\bibliographystyle{splncs04}
\bibliography{main}
%




\end{document}